\documentclass[]{piparticle-final}
\usepackage{graphicx}
\usepackage{amsmath}
\usepackage{color}
\definecolor{blue-violet}{rgb}{0.54, 0.17, 0.89}

\usepackage{cite} 
\usepackage[normalem]{ulem}  
\usepackage{color} 
\definecolor{ao}{rgb}{0.0, 0.5, 0.0}
\usepackage{epstopdf}

\begin{document}

\volume{8}               
\articlenumber{080005}   
\journalyear{2016}       
\editor{K. Hallberg}   
\reviewers{D. C. Cabra, Instituto de F\'isica La Plata, La Plata, Argentina}  
\received{1 July 2016}     
\accepted{30 August 2016}   
\runningauthor{F. T. Lisandrini \itshape{et al.}}  
\doi{080005}         

\title{Topological quantum phase transition in strongly correlated Kondo insulators in 1D}

\author{Franco T. Lisandrini,\cite{inst1,inst2}  
        Alejandro M. Lobos,\cite{inst1} 
        Ariel O. Dobry,\cite{inst1,inst2}
        Claudio J. Gazza\cite{inst1,inst2}\thanks{E-mail: gazza@ifir-conicet.gov.ar}}

\pipabstract{We investigate, by means of a field-theory analysis combined with the density-matrix renormalization group (DMRG) method, 
a theoretical model for a strongly correlated 
quantum system in one dimension realizing a topologically-ordered  Haldane phase ground state. The model consists of a spin-1/2 Heisenberg chain coupled to 
a tight-binding chain via two competing Kondo exchange couplings of different type: a ``$s$-wave'' Kondo coupling ($J^s_{K}$), and a less 
common ``$p$-wave'' ($J^p_{K}$) Kondo coupling. While the first coupling is the standard Kondo interaction studied in many condensed-matter 
systems, the latter has been recently introduced by Alexandrov and Coleman [Phys. Rev. B \textbf{90}, 115147 (2014)] as a possible mechanism 
leading to a topological Kondo-insulating ground state in one dimension. As a result of this competition, a topological quantum
phase transition (TQPT) occurs in the system for a critical value of the ratio $J^s_{K}/J^p_{K}$, separating a (Haldane-type) topological phase  
from a topologically trivial ground state where the system 
can be essentially described as a product of local singlets. We study and characterize the TQPT by means of the magnetization
profile, the entanglement entropy and the full entanglement spectrum of the ground state. Our results might be relevant to understand how 
topologically-ordered phases of fermions emerge in strongly interacting quantum systems.}
\maketitle

\blfootnote{
\begin{theaffiliation}{99}
   \institution{inst1} Instituto de F{\'i}sica Rosario (CONICET), Bv 27 de Febrero 210 bis, S2000EZP Rosario, Santa Fe, Argentina.
   \institution{inst2} Facultad de Ciencias Exactas, Ingenier{\'i}a y Agrimensura, Universidad Nacional de Rosario, Argentina.
\end{theaffiliation}}

\section{Introduction}

The study of topological quantum phases of matter has become an area of great interest in present-day condensed matter physics.
A topological phase is a quantum phase of matter which cannot be
characterized by a local order parameter, and thus falls beyond
the Landau paradigm. In particular, topological insulators (i.e., materials which are insulating in
the bulk but support topologically protected gapless states at the edges) were first proposed theoretically for two- and
three-dimensional systems
with time-reversal symmetry \cite{Bernevig06_QSHE_in_HgTe,Fu07_Topological_insulators_with_inversion_symmetry,Zhang2009}, and
soon after found in
experiments on HgTe quantum wells \cite{Konig07_Observation_of_QSH_in_HgTe} and in Bi$_{1-x}$Sb$_{x}$
\cite{Hsieh08_Topological_Dirac_insulator_in_QSH_phase}, and Bi$_{2}$Se$_{3}$ \cite{Xia09_Topological_insulator_in_Bi2Se3},
generating a lot of excitement and subsequent research.
The electronic structure of a
topological insulator cannot be smoothly connected to that of
a trivial insulator, a fact that is mathematically expressed in
the existence of a nonzero topological “invariant,” an integer
number quantifying the non-local topological order in the ground state.
A complete classification based on the dimensionality and underlying symmetries
has been achieved in the form of a ``periodic table'' of
topological insulators \cite{Altland97_Symmetry_classes,Kitaev_TI_classification,Ryu10_Topological_classification}.
Nevertheless, this classification
refers only to the gapped phases of noninteracting fermions,
and leaves open the problem of characterizing and classifying
strongly interacting topological insulators. This is a very
important open question in modern condensed-matter physics.

Recently, Dzero \textit{et al.}
\cite{Dzero10_Topological_Kondo_Insulators,Dzero12_Theory_of_topological_Kondo_insulators,Dzero15_Review_TKIs} proposed a new
kind of topological insulator: the topological Kondo insulator (TKI), which combines features of both non-interacting
topological insulators and the well-known Kondo insulators, a special class of heavy-fermion system with an insulating gap
strongly renormalized by interactions. Within a mean-field picture, TKIs can
be understood as a strongly renormalized $f$-electron band lying
close to the Fermi level, and hybridizing with the conduction-electron
$d-$bands \cite{read83,coleman87,newns87}.
At half-filling, an insulating state appears due to the opening of
a low-temperature hybridization gap at the
Fermi energy induced by interactions. Due to the opposite parities of the states being hybridized,
a topologically non-trivial ground state emerges, characterized by
an insulating gap in the bulk and conducting Dirac states at the surface \cite{Dzero10_Topological_Kondo_Insulators}.
At present, TKI materials, among which samarium hexaboride (SmB$_{6}$)
is the best known example, are under intense investigation both theoretically
and experimentally \cite{Wolgast13_SS_in_SmB6,Zhang13_SS_in_SmB6,Xu14_SARPES_in_SmB6,Kim14_Topological_SS_in_SmB6} .

From a theoretical point of view, TKIs are interesting systems arising from the interplay between strong interactions and
topology.
Although the large-$N$ meanfield approach was successful in describing qualitatively heavy-fermion systems and TKIs in
particular, it would be desirable to understand better how TKIs emerge. In order to shed more light into this question, in a recent work
Alexandrov and Coleman proposed a one-dimensional model for a topological Kondo insulator \cite{Alexandrov2014_End_states_in_1DTKI},
the ``$p$-wave'' Kondo-Heisenberg model ($p$-KHM). Such a model consists of a chain of spin-1/2 magnetic impurities interacting 
with a half-filled one-dimensional electron gas through a Kondo exchange
(see Fig. \ref{system}). 
Using a standard mean-field description \cite{read83,coleman87,newns87}, which expresses the original interacting problem as an
effectively non-interacting one, the authors mentioned above found a topologically non-trivial insulating ground state
(i.e., topological class D \cite{Altland97_Symmetry_classes,Kitaev_TI_classification,Ryu10_Topological_classification})
which hosts magnetic states at the open ends of the chain. However, this system was studied recently using the Abelian bosonization approach
combined with a perturbative renormalization group analysis, revealing
an unexpected connection to the Haldane phase at low temperatures \cite{Lobos15_1DTKI}. The Haldane phase is a paradigmatic
example
of a strongly interacting topological system
\cite{affleck_klt_short,affleck_klt_long,kennedy_z2z2_haldane, Pollmann10_Entaglement_spectrum_of_topological_phases_in_1D, Pollmann12_SPT_phases_in_1D}.
The results in Ref. \cite{Lobos15_1DTKI} indicate that
1D TKI systems might be much more complex and richer than expected
with the na\"ive mean-field approach, as they are uncapable of describing the full complexity of the Haldane phase, and suggest
that
they must be reconsidered from the more general perspective of interacting symmetry-protected topological (SPT) 
phases \cite{Gu09_SPT_Classification,Chen11_Classification_of_gapped_spin_systems_in_1D, Pollmann10_Entaglement_spectrum_of_topological_phases_in_1D, Pollmann12_SPT_phases_in_1D}. 
More recently, two numerical studies using exact DMRG methods have confirmed that 1D TKIs belong to the universality class of
Haldane
insulators \cite{Mezio15_Haldane_phase_in_1DTKIs,Legeza16_Characterization_1DTKI}. These studies have extended the regime of
validity of
the results in Ref. \cite{Lobos15_1DTKI}.

In this work, we theoretically investigate the robustness of the Haldane phase in one-dimensional topological Kondo insulators,
and study
the effect of local interactions that destabilize the topological phase and drive the system to a non-topolgical phase. Our goal is to 
characterize the system at, and near to, the
topological quantum phase transition (TQPT) from the perspective of symmetry-protected topological phases, using the concepts of entanglement entropy and
entanglement spectrum to detect the topologically-ordered ground states. This is a novel perspective in the context of TKIs, which might shed new light on the 
emergence of topological order on strongly correlated phases of fermions, and  makes our work interesting from the pedagogical and conceptual points of view.

\section{Theoretical model}

\begin{figure}
\begin{center}
\includegraphics[width=0.45\textwidth]{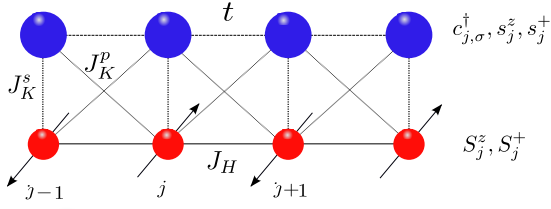}
\end{center}
\vspace{-1.0cm}
\caption{Sketch of the Kondo-Heisenberg model under consideration. The lower leg represents a spin-1/2 Heisenberg chain with $J_H>0$. The upper leg represents a 
half-filled one-dimensional tight binding chain interacting with the lower leg 
through two different Kondo exchange couplings, a ``$s$-wave'' $J_K^s$ and a ``$p$-wave'' $J_K^p$. We also show the fermionic and spin operators defined on each 
eight-dimensional ``supersite'' $j$ (see text).} \label{system}
\end{figure}

We describe the system depicted in Fig. \ref{system} with the Hamiltonian
$H=H_{1}+H_{2}+H^{\left(s\right)}_{K}+H^{\left(p\right)}_{K}$, where the
conduction band is represented by a $L$-site tight-binding chain

\begin{align}
H_{1}&=-t\sum_{j=1,\sigma}^{L-1}\left(c_{j,\sigma}^{\dagger}c_{j+1,\sigma}+\mbox{H.c.}\right)\label{H1},
\end{align}
with $c_{j,\sigma}^{\dagger}$, the creation operator of an electron
with spin $\sigma$ at site $j$. The Hamiltonian

\begin{align}
H_{2}&=J_{H}\sum_{j=1}^{L-1}\mathbf{S}_{j}\cdot\mathbf{S}_{j+1}\ \ \ (J_H>0), \label{H2}
\end{align}
corresponds to a spin-1/2 antiferromagnetic Heisenberg chain. The terms $H^{\left(s\right)}_{K}$ and $H^{\left(p\right)}_{K}$
describe two different types of Kondo exchange couplings
between $H_{1}$ and $H_{2}$, namely

\begin{align}
H^{\left(s\right)}_{K} & =J^s_{K}\sum_{j=1}^{L}\mathbf{S}_{j}\ .\ \mathbf{s}_{j},\label{eq:H_sK}\\
H^{\left(p\right)}_{K} & =J^p_{K}\sum_{j=1}^{L}\mathbf{S}_{j}\ .\ \boldsymbol{\pi}_{j},\label{eq:H_pK}
\end{align}
with $J^a_{K}>0$ ($a=s,p$). Eq. (\ref{eq:H_sK}) describes the usual antiferromagnetic Kondo exchange coupling of a spin
$\mathbf{S}_{j}$ in the Heisenberg chain to the local spin density in the fermionic chain at site $j$, defined as:

\begin{align}
\mathbf{s}_{j}&\equiv\sum_{\alpha,\beta}c_{j,\alpha}^{\dagger}\bigl(\frac{\boldsymbol{\sigma}_{\alpha\beta}}{2}\bigr)c_{j,\beta},\label{eq:sj}
\end{align}
where $\boldsymbol{\sigma}_{\alpha\beta}$ is the vector of Pauli matrices.
On the other hand, Eq. (\ref{eq:H_pK}) describes a ``\textit{p-}wave'' Kondo interaction, which is unusual in that it couples
the spin $\mathbf{S}_{j}$ to the \textit{p-}wave spin density in the fermionic chain at site $j$, defined as:
\cite{Alexandrov2014_End_states_in_1DTKI}

\begin{align}
\boldsymbol{\pi}_{j}&\equiv \label{eq:pij} \\ 
&\sum_{\alpha,\beta}\bigl(\frac{c_{j+1,\alpha}^{\dagger}-c_{j-1,\alpha}^{\dagger}}{\sqrt{2}}\bigr)\bigl(\frac{\boldsymbol{\sigma}_{\alpha\beta}}{2}\bigr)\bigl(\frac{c_{j+1,\beta}-c_{j-1,\beta}}{\sqrt{2}}\bigr),   \notag 
\end{align}
where the notation $c_{0,\sigma}=c_{L+1,\sigma}=0$ is implied.

The case $J^s_K>0$ and $J^p_K=0$
corresponds to the so-called ``Kondo-Heisenberg model'' in 1D, which has been extensively studied in the past in connection to
the stripe phase of high-$T_c$ superconductors
\cite{zachar_Kondo_chain_toulouse,Sikkema97_Spin_gap_in_a_doped_Kondo_chain,zachar_exotic_Kondo,Zachar01_Staggered_phases_1D_Kondo_Heisenberg_model,Berg10_Pair_density_wave_in_Kondo_Heisenberg_Model,Dobry13_SC_phases_in_the_KH_model,Cho14_Topological_PDW_superconducting_state_in_1D}.
These previous works indicate that, at half-filling, that model does not support any topological phases. On the other hand, the
case $J^s_K=0$ and $J^p_K>0$ is the ``$p$-wave'' Kondo-Heisenberg model proposed recently in Ref. \cite{Alexandrov2014_End_states_in_1DTKI}, and subsequently studied in Refs.
\cite{Lobos15_1DTKI,Mezio15_Haldane_phase_in_1DTKIs,Legeza16_Characterization_1DTKI}, where it was stablished that the
ground state corresponds to the Haldane phase. In both cases, at half-filling the system develops a Mott-insulating gap in the
fermionic chain due to ``umklapp'' processes (i.e., backscattering processes with $2k_F$ momentum transfer) originated in the
Kondo interactions $J^s_K$ or $J^p_K$, and at low temperatures (lower than the Mott gap), the system effectively maps onto a
spin-1/2 ladder. However, the ground states of the model in both cases cannot be smoothly connected (i.e., one is
topologically trivial while the other is not), and therefore we anticipate that a gap-closing topological quantum phase transition
(TQPT) must occur as a function of the ratio $J^s_K/J^p_K$.
Indeed, in Ref. \cite{Lobos15_1DTKI}, it was proposed that while in the first case the (effective) spin-1/2 ladder forms
\textit{singlets} along the rungs, in the second case the Kondo coupling $J^p_K$ favors the formation of \textit{triplets} on
the rungs, and therefore the system maps onto the Haldane spin-1 chain
\cite{shelton_spin_ladders,gogolin_disordered_ladder,lecheminant02_magnet,
Robinson12_Finite_wavevector_pairing_in_doped_ladders}. From this perspective, our toy-model Hamiltonian $H$ allows to explore a
transtition from topological to non-topological Kondo insulators, and to gain a valuable insight of the TQPT.

\section{Field-theory analysis}

The purpose of this section is to provide a simple and phenomenological understanding of the competition between the two Kondo interactions $J_K^s$ and $J_K^p$, which is not evident in Eqs. (\ref{eq:H_sK}) and (\ref{eq:H_pK}).
To that end, we introduce a field-theoretical representation of the
model, valid at sufficiently low temperatures. We linearize the non-interacting spectrum
$\epsilon_{k}=-2t\cos\left(ka\right)$
in the tight-binding  chain $H_{1}$ around the Fermi energy $\mu=0$,
and take the continuum limit $a\rightarrow0$, where $a$ is the lattice constant.
Then, the low-energy representation of the fermionic operators becomes
 \cite{Emery_1D, Solyom79, vondelft_bosonization_review, giamarchi_book_1d,gogolin_1dbook}

\begin{align}
\frac{c_{j,\sigma}}{\sqrt{a}} & \sim
e^{ik_{F}x_{j}}R_{1,\sigma}\left(x_{j}\right)+e^{-ik_{F}x_{j}}L_{1,\sigma}\left(x_{j}\right),\label{eq:linearization}
\end{align}
where $R_{1,\sigma}\left(x\right)$ and $L_{1,\sigma}\left(x\right)$
are right- and left-moving fermionic field operators, which vary slowly
on the scale of $a$. Using this representation, the spin densities become  \cite{affleck_houches}:

\begin{align}
\frac{\mathbf{s}_{j}}{a} &\rightarrow
\left[\mathbf{J}_{1R}\left(x_{j}\right)+\mathbf{J}_{1L}(x_{j})+\left(-1\right)^{j}\mathbf{N}_{1}\left(x_{j}\right)\right],\label{eq:sj_density}\\
\frac{\boldsymbol{\pi}_{j}}{a} &\rightarrow 2 \left[\mathbf{J}_{1R}\left(x_{j}\right)+\mathbf{J}_{1L}\left(x_{j}\right)\right. \notag \\
 & \ \ \ \ \ \ \ \left.-(-1)^{j}\mathbf{N}_{1}\left(x_{j}\right)\right],\label{eq:pi_density}
\end{align}
where we have defined slowly-varying spin densities for the $smooth$  spin configurations:

\begin{align}
\mathbf{J}_{1R}\left(x_{j}\right)&=\sum_{\alpha,\beta} R^\dagger_{1,\alpha}\left(x_{j}\right)
\Bigl(\frac{\boldsymbol{\sigma}_{\alpha\beta}}{2}\Bigr)R_{1,\beta}\left(x_{j}\right),\\
\mathbf{J}_{1L}\left(x_{j}\right)&=\sum_{\alpha,\beta} L^\dagger_{1,\alpha}\left(x_{j}\right)
\Bigl(\frac{\boldsymbol{\sigma}_{\alpha\beta}}{2}\Bigr)L_{1,\beta}\left(x_{j}\right),
\end{align}
and for the $staggered$ spin density

\begin{align}
\mathbf{N}_{1}\left(x_{j}\right)&=\sum_{\alpha,\beta} R^\dagger_{1,\alpha}\left(x_{j}\right)
\Bigl(\frac{\boldsymbol{\sigma}_{\alpha\beta}}{2}\Bigr)L_{1,\beta}\left(x_{j}\right) + \text{H.c.}.
\end{align}

Similarly, a continuum representation for the Heisenberg chain 
can be achieved, e.g., by fermionization of the $S$=1/2 spins by means of a Jordan-Wigner transformation. At
low energies, the spin densities become \cite{affleck_houches, shelton_spin_ladders,gogolin_disordered_ladder,lecheminant02_magnet,
Robinson12_Finite_wavevector_pairing_in_doped_ladders}

\begin{align}
\frac{\mathbf{S}_{j}}{a} &
\rightarrow\mathbf{J}_{2R}\left(x_{j}\right)+\mathbf{J}_{2L}\left(x_{j}\right)+\left(-1\right)^{j}\mathbf{N}_{2}\left(x_{j}\right).\label{eq:S_density}
\end{align}

We now focus on the Kondo interaction and leave the analysis of  $H_1$ and $H_2$ aside, as these terms are unimportant for the qualitative understading of the basic
mechanism leading to the TQPT. Keeping the most relevant  (in
the RG sense) terms, we can write the Kondo interaction as:

\begin{align}
H^{\left(s\right)}_{K}+H^{\left(p\right)}_{K} & \rightarrow \left(J^s_{K}-2 J^p_{K}\right)\int dx\ \mathbf{N}_{1}\left(x
\right) . \mathbf{N}_{2}\left(x \right) \nonumber\\
&(+ \text{less relevant contributions}),\label{eq:HK_continuum}
\end{align}
i.e., the Kondo interaction couples the staggered magnetization components in chains 1 and 2. In the above expression, note that while a large $J_K^s$ favors a positive value of the effective coupling $(J^s_{K}-2 J^p_{K})$, therefore promoting the formation of local singlets along the rungs, a large $p$-wave Kondo coupling $J_K^p$ favors  an 
{\it effective ferromagnetic} coupling which promotes the formation of local triplets with $S=1$ (hence the connection to the $S=1$ Haldane chain). The minus sign in front of $J^p_{K}$ appears as a result of the $p$-wave nature of the orbitals in
Eq. (\ref{eq:pij}). From this qualitative analysis we can conclude that $J_K^s$ and
$J_K^p$ will be competing interactions promoting different ground states, and since these grounstates cannot be adiabatically connected with each other, a TQPT must occur. 

Strictly speaking, near the critical region where the bare coupling $ \left(J^s_{K}-2 J^p_{K}\right)$ vanishes, the less relevant terms neglected  in Eq. (\ref{eq:HK_continuum}) should be taken into account. However, note that operators with conformal spin 1 (i.e., operators of the form $ \left(\partial_x\mathbf{N}_{1}\right) .  \mathbf{N}_{2}$ or  $ \mathbf{N}_{1}. \left(\partial_x \mathbf{N}_{2}\right)$, see for instance Ref. \cite{nersesyan_incom}) are not allowed by the inversion symmetry of the Hamiltonian and the $p-$wave symmetry of the orbitals in Eq. (\ref{eq:pij}), which demands that $c_{j+1,\alpha}-c_{j-1,\alpha}\rightarrow-\left(c_{-(j+1),\alpha}-c_{-(j-1),\alpha}\right)$ under the change $j\rightarrow -j $, and therefore forbids the occurrence of terms proportional to $\partial_x \mathbf{N}_1$. Therefore, only the marginal operators $ \mathbf{J}_{1 \nu}. \mathbf{J}_{2 \nu^\prime}$ (with $\nu=\left\{R,L\right\}$) and terms with conformal spin bigger than 1 are expected in the Hamiltonian. We do not expect these operators to  change the physics qualitatively near the critical point, and we can ignore them for this simplified analysis. 
As we show below, our numerical DMRG results are in accordance with this qualitative picture.

\section{DMRG analysis}\label{dmrg}

Before presenting the numerical results, it is worth providing technical details on the implementation of the DMRG method applied
to the present model. This particular Hamiltonian contains two types of terms: (a) terms involving two local operators, as in most 
condensed-matter models with nearest-neighbor interactions, i.e., Eqs. (\ref{H1})-(\ref{eq:H_sK}), and 
(b) terms involving three local operators, which result from the expansion of Eq. (\ref{eq:H_pK}). 
To make it easier to implement, we have found useful to define first a ``supersite'' representation of the system, where each 
supersite combines a spin $\mathbf{S}_j$ and the fermionic site along each rung (see Fig. \ref{system}), therefore spanning a new 
8-dimensional local basis.
The first kind of terms could be easily handled with standard DMRG implementations where the system is represented as 
$L(j)\otimes \bullet \otimes\bullet \otimes R(N-j-2)$, with $L(j)$ and $R(j)$ the left and right blocks with $j$ supersites, respectively, 
and the two circles are the exactly-represented middle supersites $j+1$ and $j+2$. 
This comes at a price, however, since one is then forced to re-express the electron creation operator $c_{j,\sigma}^{\dagger}$, 
and the spin-1/2 operator $\mathbf{S}_{j}$ in this new basis in order to implement Eqs. (\ref{H1}) and (\ref{H2}).
In this basis, note that the Hamiltonian $H^{\left(s\right)}_{K}$, Eq. (\ref{eq:H_sK}), becomes an ``on-site'' term, which can be handled easily. 

In the second type of contributions, the presence of three-operator terms in $H^{\left(p\right)}_{K}$, Eq. (\ref{eq:H_pK}), must be properly treated in order to avoid
extra truncation errors due to the tensor product of two (already truncated) operators inside each left and right blocks. 
Then, during each left-right DMRG sweep iteration, we save in the 
previous superblock configuration $L(j-1)\otimes \bullet \otimes\bullet \otimes R(N-j-3)$ the
exact correlation matrices between spin and fermionic operators that involve positions $j-1$ (i.e., rightmost supersite of left
block) and $j$ (first single site), which will become the two rightmost supersites of the new $L(j)$ in the next sweep iteration step.
These correlation matrices are

\begin{align}
\left [ A_{j-1,j,\sigma}^{z \dagger}\right ]_{i;i' }&= \notag \\
 &\rho_{i;i1i2} \left [ S_{j-1}^{z}\right ]_{i1;i1'} \left [
c_{j,\sigma}^{\dagger} \right ]_{i2;i2' }\rho_{i1'i2';i'}^{\dagger},\notag 
\end{align}
and

\begin{align}
\left [ A_{j-1,j,\overline{\sigma}}^{+ \dagger}\right ]_{i;i' }&= \notag\\
&\rho_{i;i1i2} \left [ S_{j-1}^{+}\right ]_{i1;i1'} \left [
c_{j,\overline{\sigma}}^{\dagger} \right ]_{i2;i2' }\rho_{i1'i2';i'}^{\dagger},\notag 
\end{align}
where $\rho_{i;i1i2}$ is the $m\times 8m$ reduced density matrix,  with $m$ the number of states kept, and where the indices $i$ and $i1$ run over the truncated $m$-dimensional Hilbert space while $i2$ runs over the 8-dimensional supersite space. In addition, we have assumed summation over repeated internal indices. 
Similarly, correlations between spin and fermionic operators placed in the two leftmost supersites $j+3$ and $j+4$ of $R(N-j-2)$ should also be
kept during the corresponding step of the rightf-left sweep. 
Therefore, we now deal with a standard ``two-operator'' interaction, for example the term
$L(j) \otimes \bullet$ of $H^{\left(p\right)}_{K}$ is

$$
-\frac{1}{4}J_K^p \sum_{\sigma } \left \{ \left (\sigma A_{j-1,j,\sigma}^{z \dagger} + 
A_{j-1,j,\overline{\sigma}}^{+ \dagger}  \right ) c_{l_{j+1},\sigma } + \text{h.c.}\right \}.
$$

Finally, we mention that in our implementation we have kept up to a maximum of 800 states and we have swept 12
times, assuring truncation errors in the density matrix of the order $10^{-9}$ at worst. 

We now turn to the results. We have used the tight-binding parameter $t$ as our unit of energies, and in all of our calculations we have used the values $J_H/t=1$ and $J_K^p/t=2$. We have studied the 
evolution of the ground state upon the increase of $J_K^s$ starting from the value $J_K^s=0$, where the system is in the Haldane phase.
The nature of the topologically-ordered  ground state and the precise detection of the TQPT are determined using, respectively: a) the analysis of the spin profile in the ground state, b) the value of the
entanglement entropy  and c) the analysis of the degeneracies in the full entanglement spectrum. 
As shown recently in seminal works \cite{Pollmann10_Entaglement_spectrum_of_topological_phases_in_1D, Pollmann12_SPT_phases_in_1D}, 
the last two properties are useful \textit{bona fide} indicators of symmetry-protected topological orders. 

\begin{figure}
\includegraphics[width=0.5\textwidth]{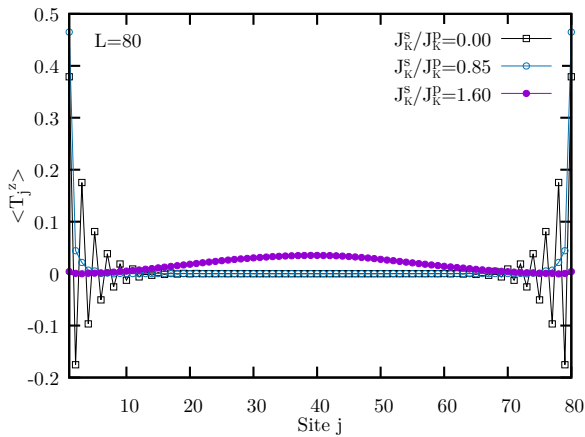}\protect\caption{\label{figure2}
 Spatial profile of   $\langle T^z_j \rangle=\langle \psi_{g}^{M^z=1}|T_{j}^{z}|\psi_{g}^{M^{z}=1}\rangle$, i.e., 
the $z$ component of the total spin in the supersite $j$, computed with the ground state of the subspace with total
$M^z=1$ for $L=80$, and for parameters $J_H/t=1$ and $J_K^p/t=2$. 
The destruction of the topologically protected spin-1/2 states at the ends of the chain can be clearly seen as $J_K^s/J_K^p$ is increased from $J_K^s/J_K^p=0$ to $J_K^s/J_K^p=1.6$.}
\end{figure}

\textit{(a) Spin profile.} One characteristic feature of a $S=1$ Haldane chain with open boundary conditions is the presence of 
topologically protected fractionalized spin-1/2 end-states, a consequence of the broken $Z_2\times Z_2$ symmetry of the 
ground state \cite{kennedy_z2z2_haldane}. These states can be represented as: $|\uparrow_L\rangle \otimes|\uparrow_R\rangle$, $|\uparrow_L\rangle\otimes|\downarrow_R\rangle$, 
$|\downarrow_L\rangle\otimes|\uparrow_R\rangle$, $|\downarrow_L\rangle\otimes|\downarrow_R\rangle$, and correspond to the fourfold-degenerate ground state in the thermodynamic limit $L\rightarrow \infty$.
In order to detect these fractionalized spin excitations in our system, we have defined the spin profile as 
$\left\langle T_{j}^{z}\right\rangle =\left\langle \psi_{g}^{M^z=1}\right|T_{j}^{z}\left|\psi_{g}^{M^z=1}\right\rangle $,
where $T_{j}^{z}$ is the $z-$projection of the total spin in the $j$-th rung $\mathbf{T}_{j}=\mathbf{S}_{j}+\mathbf{s}_{j}$,
and $\left|\psi_{g}^{M^z=1}\right\rangle $ is the ground state of the system with total spin $M^z=1$ (where the visualization of the spin states at the ends is easier). 
In Fig. \ref{figure2} we show $\langle T^z_{j}\rangle$ vs $j$ for different values of $J_K^s/J_K^p$ and for $L=80$.
The presence of spin states localized at the edges can be clearly seen for  $J_K^s/J_K^p=0$ and $J_K^s/J_K^p=0.85$, where the spin density is accumulated at the ends of the system. Note that since we are working in the subspace $M^z=1$, this ground state corresponds to the state $|\uparrow_L\rangle \otimes|\uparrow_R\rangle$.
For $J_K^s/J_K^p=1.6$, the magnetic edge-states have already disappeared, indicating that the onset of the topologically trivial phase must occur at lower values of $J_K^s/J_K^p$. However, while this analysis is useful to understand the nature of the topologically-ordered ground state, it does not allow a precise determination of the TQPT. To that end, we 
have studied the entanglement entropy and entanglement spectrum (see below).

\begin{figure}
\begin{center}
\includegraphics[width=0.5\textwidth]{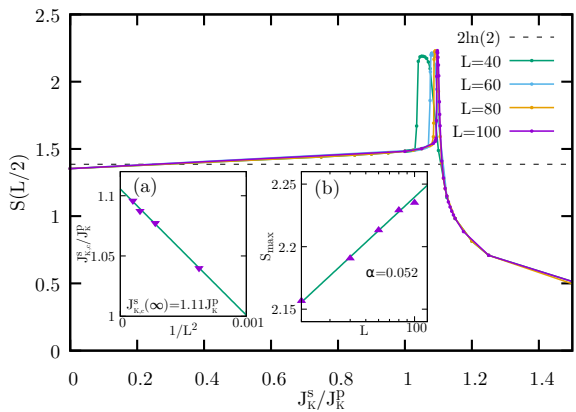}
\end{center}
\caption{Entanglement entropy of the reduced density matrix, $S\left(L/2\right)$,  as function of $J_K^s/J_K^p$ for different lattice sizes. The entropy is computed within  the subspace $M^z=0$ and for parameters $J_H/t=1$ and $J_K^p/t=2$. The maximum of  $S\left(L/2\right)$ indicates the position of the critical point. Inset (a): Finite-size scaling  of the critical point, i.e., $J_{K,c}^s\left(L \right)=J_{K,c}^s\left(\infty \right) + a /L^2$, from where the value in the thermodynamic limit $J_{K,c}^s\left(\infty \right)\approx 1.11\,J_K^p$ is obtained. Inset (b): Maximum entropy $S_\text{max}(L/2)$ vs $L$. The results reproduce the predicted logarithmic divergence $S_\text{max}(L/2)=\alpha \ln(L) + \text{constant}$, with the fitting constant $\alpha=0.052$.
} \label{entanglement_entropy}
\end{figure}

\textit{(b) Entanglement entropy.} We have also calculated the entanglement entropy (i.e., the von Neumann entropy of the
reduced density matrix), defined as \cite{Schollwoeck05_Review_DMRG}

\begin{align}
S(L/2)&=-\text{Tr}\hat{\rho}_{L/2}\ln \hat{\rho}_{L/2}=-\sum_{j}\Lambda_{j}\ln \Lambda_{j},
\end{align}
where $\hat{\rho}_{L/2}$ is the reduced density matrix obtained after tracing out half of the chain, and $\Lambda_{j}$ the
corresponding eigenvalues of $\hat{\rho}_{L/2}$, which are the squares of the Schmidt values. Recently, it has been clarified that the entanglement of a single quantum
state is a crucial property not only from the perspective of quantum information, but also for condensed matter physics. In
particular, the entanglement entropy has been shown to contain the \textit{quantum dimension}, a property of topologically-ordered
phases \cite{Kitaev06_Topological_entanglement_entropy, Levin06_Topological_entanglement_entropy}. Hirano and 
Hatsugai \cite{Hirano07_Entanglement_entropy_of_gapped_spin_chains} have computed the entanglement entropy of the open-boundary spin-1 
Haldane chain and obtained the lower-bound value  $S(L/2)= \ln\left(4\right)=2\ln\left(2\right)$ which, according to the edge-state 
picture in the thermodynamical limit $L\rightarrow \infty$, corresponds to the aforementioned 4 spin-1/2 edge
states. In
Fig. \ref{entanglement_entropy} we show the entanglement entropy of the system as a function of  $J_K^s/J_K^p$, for different system sizes and in the subspace $M^z=0$, where we expect to find the ground state (i.e., the ground state of an  even-numbered antiferromagnetic chain is a global singlet \cite{auerbach_book_spins}). Near the critical region,   
 the entanglement entropy grows due to the contribution of the bulk, and exactly at the TQPT the entanglement entropy is predicted to show a logarithmic divergence $S(L/2)\sim \ln(L)$, characteristic of critical one-dimensional systems \cite{Schollwoeck05_Review_DMRG}. 
 As the size of the system is increased, the logarithmic divergence becomes narrower and its position shifts to larger values. Using the fitting function  $J_{K,c}^s\left(L \right)=J_{K,c}^s\left(\infty \right) + a /L^2$, we have obtained the extrapolated critical point $J_{K,c}^s\left(\infty \right)\approx 1.11\,J_K^p$ in the thermodynamic limit, see inset (a) in Fig. \ref{entanglement_entropy}. Note that this value is smaller than the predicted value $J_{K,c}^s= 2\,J_K^p$ using the field-theory analysis of the previous section. We believe this to be the effect of the neglected marginal or irrelevant operators, which renormalize non-universal quantities such as the critical point.

We have also confirmed the logarithmic scaling of the entanglement entropy at the critical point, i.e., $S_\text{max}(L/2)=\alpha \ln(L) + \text{constant}$, and we have obtained a prefactor $\alpha=0.052$, see inset (b) in Fig. \ref{entanglement_entropy}. A detailed analysis of this value and its connection to the corresponding central charge value   of the conformal field theory is  beyond the scope of the present work and is left to a subsequent publication.
 
In Fig. \ref{entanglement_entropy}, note that  for  $ J_K^s/J_K^p< J_{K,c}^s/J_K^p$,  the value of the entanglement entropy  roughly corresponds to 
 $S(L/2)\sim 2\ln\left(2\right)$, consistent with the theoretical predictions in the Haldane phase. Values of $S(L/2)$ which are below the predicted lower-bound are presumably due to finite-size effects, which result in a decrease of the effective quantum dimension 
 in small systems \cite{Hirano07_Entanglement_entropy_of_gapped_spin_chains}.  For $ J_K^s/J_K^p> J_{K,c}^s/J_K^p$, $S(L/2)$ tends to zero as expected for a topologically trivial ground state. This result can be  easily understood in the limit $J_K^s/J_K^p \rightarrow \infty$, where we expect the ground state to factorize as a product of 
 local singlets (we recall that we are working in the subspace $M^z=0$), for which $S(L/2)=0$. 

\begin{figure}
\begin{center}
\includegraphics[width=0.5\textwidth]{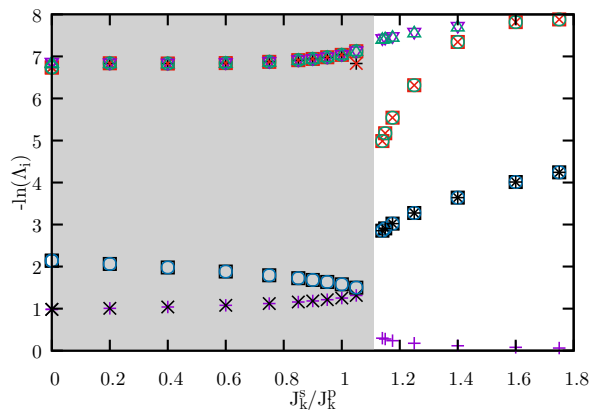}
\end{center}
\caption{The largest eigenvalues of the entanglement spectrum (extrapolated to the thermodynamic limit $L\rightarrow \infty$), as a function of $J_K^s$. The results were obtained in the subspace $M^z=0$. 
The gray zone corresponds to the Haldane phase, where the degeneracy in the spectrum of eigenvalues is even.} \label{entanglement_spectrum}
\end{figure}

\textit{(c) Degeneracy of the entanglement spectrum.} Finally, we focus on the full entanglement spectrum of the reduced density-matrix. Degeneracies in the entanglement spectrum are intimately related to the
existence of discrete symmetries which protect the topological order. In particular, as shown in Ref. \cite{Pollmann10_Entaglement_spectrum_of_topological_phases_in_1D},  an even degeneracy  constitutes the most distinctive feature of the Haldane phase. This fact allows to make an interesting connection 
between the theory of symmetry-protected topological phases and the theory of topological Kondo insulators in one dimension, and might have important implications in the understanding of strongly interacting topological phases of fermions. In Fig. \ref{entanglement_spectrum} we show the evolution 
of the largest eigenvalues extrapolated to the thermodynamic limit of the density matrix as a function of the parameter $J_K^s/J_K^p$, obtained in the subspace $M^z=0$. 
Note that for  $J_K^s/J_K^p< J_{K,c}^s/J_K^p \sim 1.11$ (i.e., gray-shaded area), the degeneracy of the eigenvalues is even, as expected for the Haldane phase. In contrast, for values $J_K^s/J_K^p > J_{K,c}^s/J_K^p$, the even-degeneracy breaks down, indicating the onset of the trivial phase. In particular, note the evolution of the largest eigenvalue of the density matrix $\Lambda_0\approx 1$ (i.e., lowest ``\textcolor{blue-violet}{\textbf{+}}'' symbol) which becomes non-degenerate.

\section{Conclusions}
Using a combination of techniques, i.e., a field-theoretical analysis and the density-matrix renormalization group (DMRG), an essentially exact method in one dimension, we have studied the transition from a topological to a non-topological phase 
in a model for a one-dimensional strongly interacting topological Kondo insulator. As a prototypical topological quantum phase transition with no Landau-type local order-parameter, one must resort to global quantities characterizing the ground state. In this work, we have shown that the 
entanglement entropy and the entanglement spectrum can be used to characterize a topological Kondo insulator in one dimension. This system was 
originally understood and classified according to  non-interacting topological invariants (i.e., Chern numbers), employing approximate large-$N$ 
mean-field methods \cite{Alexandrov2014_End_states_in_1DTKI}. Here, by the means of the DMRG, we have shown that a more appropriate way to 
understand this system is by using the concepts developed for symmetry-protected topological 
phases \cite{Gu09_SPT_Classification,Chen11_Classification_of_gapped_spin_systems_in_1D, Pollmann10_Entaglement_spectrum_of_topological_phases_in_1D, Pollmann12_SPT_phases_in_1D}. In particular, for parameters $J_H/t=1$ and $J_K^p/t=2$, we have obtained the value of the critical point $J_{K,c}^s/J_K^p\simeq 1.11$ in the thermodynamic limit $L\rightarrow \infty$. This value is smaller than the expected from the qualitative field-theoretical estimation $J_{K,c}^s/J_K^p = 2$, a fact that is presumably originated in the effect of  marginal or irrelevant operators, which were neglected in the qualitative analysis and which renormalize a non-universal quantity such as the critical point.


\begin{acknowledgements}
F.T.L, A.O.D. and C.J.G. acknowledge support from CONICET-PIP 11220120100389CO. A.M.L acknowledges support from  PICT-2015-0217 of ANPCyT.
\end{acknowledgements}

\bibliographystyle{unsrt}

\end{document}